\documentclass{aa}
\usepackage{graphicx}
\usepackage{txfonts}
\begin{document}
   \title{Intrinsic absorbers in BL Lac objects: the XMM-Newton view}

   \author{A. J. Blustin
          \and
          M. J. Page
          \and
          G. Branduardi-Raymont
          }
   \offprints{A. J. Blustin\\
             \email{ajb@mssl.ucl.ac.uk}}

   \institute{MSSL, University College London,
             Holmbury St. Mary, Dorking, Surrey RH5 6NT, England
	  }

   \date{Received 16 October 2003 / Accepted 15 December 2003}

   \abstract{We present XMM-Newton observations of four BL Lac objects (\object{1H1219+301}, \object{H1426+428}, \object{Markarian 501} and \object{PKS 0548-322)}, which have been found with past X-ray missions to contain evidence of broad soft X-ray absorption features. Observations with the high resolution Reflection Grating Spectrometers on XMM-Newton provide the best chance yet of investigating these features. No broad absorption features are observed in any of the objects. Neither do we find convincing evidence for narrow emission and absorption lines in the RGS spectra. We discuss the history of observations of broad absorption features in these four objects, finding that if the features exist then they must be transient, and that - given the frequency of reported observations of them - we can rule out the existence of transient broad absorption features in these objects at 93\% confidence.
   }

   \maketitle
%

\section{Introduction}

BL Lacs are well-known for their featureless spectra, all the way from radio to gamma rays. These smooth continua are thought to be generated by synchrotron self-compton emission from jets of material streaming out of the active nucleus towards us. Canizares \& Kruper (\cite{canizares}), however, discovered a soft X-ray absorption feature in the Einstein Objective Grating Spectrometer (OGS) spectrum of the BL Lac PKS~2155-304, and since that time, similar features have been detected in several other objects: 1H1219+301, Markarian~501, PKS~0548-322, Markarian~421 (Madejski et al. \cite{madejski1991}) and H1426+428 (Madejski et al. \cite{madejski1992}). This absorption, interpreted in each case as being due to ionised gas intrinsic to the BL Lac, is perhaps the only insight we have into the immediate environment of these enigmatic sources. The dispersive spectrometers on-board XMM-Newton and Chandra, with their combination of high resolution and large effective area, provide the best opportunity yet to investigate and understand these features. To this end, XMM-Newton observed the BL Lacs 1H1219+301, H1426+428, Markarian~501 and PKS~0548-322 with the aim of obtaining high resolution soft X-ray spectra with the Reflection Grating Spectrometers (RGS; den Herder et al \cite{denherder}), and high signal-to-noise CCD resolution spectra from EPIC (Str\"{u}der et al. \cite{struder}). This paper presents the X-ray spectra of these objects, and discusses the evidence for spectral traces of the BL Lac environment. Table~\ref{source_properties} gives their coordinates, cosmological redshifts and Galactic neutral absorbing columns.

   \begin{table*}
    
      \caption[]{The coordinates, redshifts and Galactic neutral Hydrogen columns of the sources.}
         \label{source_properties}  
   \centering
         \begin{tabular}{p{1in}p{1.1in}p{1.1in}p{0.8in}p{1in}}
            \hline
            \noalign{\smallskip}
            Target       & Right Ascension & Declination     & Redshift$^{\mathrm{a}}$ & Galactic column$^{\mathrm{b}}$ \\
            \noalign{\smallskip}
            \hline
            \noalign{\smallskip}
            1H1219+301    & 12h 21m 21.941s & +30d 10m 37.11s & 0.182                   & 1.78$^{\mathrm{c}}$  \\
            H1426+428     & 14h 28m 32.6s   & +42d 40m 21s    & 0.129                   & 1.36$^{\mathrm{d}}$ \\
            Markarian~501 & 16h 53m 52.2s   & +39d 45m 37s    & 0.03366                 & 1.73$^{\mathrm{c}}$  \\
            PKS~0548-322  & 5h 50m 40.771s  & -32d 16m 17.76s & 0.069                   & 2.51$^{\mathrm{d}}$  \\
            \noalign{\smallskip}
            \hline
         \end{tabular}
\begin{list}{}{}
\item[$^{\mathrm{a}}$] redshift as listed by the NASA Extragalactic Database (NED)
\item[$^{\mathrm{b}}$] Galactic absorbing column in 10$^{\rm 20}$ cm$^{\rm -2}$
\item[$^{\mathrm{c}}$] Elvis et al. \cite{elvis}
\item[$^{\mathrm{d}}$] Costamante et al. \cite{costamante}
\end{list}
  \end{table*}

Various authors have searched for soft X-ray ionised absorption in the four BL Lacs discussed in this paper. Madejski et al. (\cite{madejski1991}) found evidence for an absorption notch (trough) in the Einstein SSS spectrum of 1H1219+301; this feature appeared at 0.56 $^{\rm +0.01}_{\rm -0.02}$~keV (all energies in this Section are quoted in the observed frame unless otherwise stated) and had a width of about 100~eV. It has apparently not been seen since with other missions. 

A soft X-ray absorption feature in the BBXRT spectrum of H1426+428 was reported by Madejski et al. (\cite{madejski1992}), and analysed in greater detail by Sambruna et al. (\cite{sambruna1997_h1}). These latter authors also find indications of soft X-ray ionised absorption in spectra from ROSAT and ASCA. The BBXRT feature was best described by a broad gaussian absorption line at $\sim$0.66~keV with a width less than or equal to 43000~km s$^{\rm -1}$. The ROSAT and ASCA data were fitted together as they had similar flux levels. The fitting produced indications of two absorption edges at 0.15~keV and 0.52~keV.

The first evidence for an absorption feature in Markarian~501 came from the Einstein SSS spectrum; Urry et al. (\cite{urry}) found an absorption trough at $\sim$0.57~keV with a width of 80~eV (these were the same as the parameters obtained for PKS~2155-304). The data were reanalysed by Madejski et al. (\cite{madejski1991}), who modelled the absoption with a notch at 0.56 $\pm$ 0.01~keV, of width 240~eV and covering factor 0.52. A later study by Fink et al. (\cite{fink}) of ROSAT data revealed the possibility of an absorption line at 0.37 $^{\rm +0.25}_{\rm -0.09}$~keV (the width was fixed at 100~eV), although the statistical significance was too low to prove the existence of the line.

Madejski et al. (\cite{madejski1991}) detected an absorption trough at 0.57 $\pm$ 0.02~keV in the Einstein SSS spectrum of PKS~0548-322. More recently, Sambruna et al. (\cite{sambruna1998_pks0548}) found an absorption edge at $\sim$0.66~keV in ASCA data; this feature could alternatively be explained by a notch at 0.82~keV with width 100~eV (fixed) and covering factor 0.12. They re-analysed the Einstein SSS data, obtaining an edge at 0.487~keV, and found evidence of two edges in ROSAT data. One of the ROSAT edges (at around 1~keV) was probably instrumental, and the other, at 0.553~keV, may either have been intrinsic or due to residual calibration problems.

The usual interpretation advanced for these features is that they originate in highly ionised oxygen in the environment of the BL Lac. The earliest detected feature amongst the objects described here, the absorption trough in Markarian~501, was simply described as being due to highly ionised oxygen (Urry et al. \cite{urry}). Madejski et al. (\cite{madejski1991}), following the lead of Canizares \& Kruper (\cite{canizares}), interpreted their notches as being due to \ion{O}{viii} Ly$\alpha$ resonance absorption. Sambruna et al. (\cite{sambruna1997_h1}) found their $\sim$0.66~keV feature to be consistent with absorption edges of \ion{O}{v}--\ion{O}{vii}, and marginally consistent with \ion{O}{viii} Ly$\alpha$. In their ROSAT and ASCA fits, no physical meaning could be determined for the 0.15~keV edge, whereas that at 0.52~keV was identified with the K-edge of \ion{O}{iii}. Fink et al. (\cite{fink}) give no physical interpretation for their 0.37~keV line. The absorption feature seen by Sambruna et al. (\cite{sambruna1998_pks0548}) in PKS~0548-322 has two possible explanations: if it is an edge, it corresponds to \ion{O}{vi}, and if it is a notch it would correspond to \ion{Fe}{xix}--\ion{Fe}{xx} or \ion{Ne}{vii}. Sambruna et al. (\cite{sambruna1997_h1} and \cite{sambruna1998_pks0548}) also fit the absorption with photoionised warm absorber models. The appearance of features at different energies, when the sources are at different flux levels, has also been used as a diagnostic. In H1426+428, the features were found to be at lower energies when the source flux was lower, consistent with the predictions of warm absorber models (Sambruna et al. \cite{sambruna1997_h1}). The opposite has been found to be true (lower energy features at \emph{higher} flux levels), however, in PKS~0548-322 (Sambruna et al. \cite{sambruna1998_pks0548}). 

Seyfert galaxies have the best understood warm absorbers of all the classes of AGN. Since the advent of high resolution X-ray spectrometers, it has been known that narrow absorption and emission lines, rather than deep absorption edges, are the most common characteristic of ionised gas in the environment of this variety of AGN (e.g. Kaastra et al. \cite{kaastra2000}). It would be interesting to see if this is also true of BL Lacs, and particularly whether the low resolution of previous spectrometers has been hiding a wealth of information in these narrow lines. Since the launch of Chandra, weak intrinsic narrow absorption lines of \ion{O}{vii} He$\alpha$ have been found in Chandra LETGS spectra of Markarian~421 (Nicastro et al. \cite{nicastro2000}, \cite{nicastro2003}), and narrow absorption lines due to the local Warm-Hot Intergalactic Medium have also been found in PKS~2155-304 (Nicastro et al. \cite{nicastro2002}) and Markarian~421 (Nicastro et al. \cite{nicastro2003}) as observed by Chandra, and also in XMM-Newton RGS spectra (Paerels et al. \cite{paerels}). There have been no reports so far of the broad absorption features found in the past, even though PKS~2155-304 and Markarian~421 have been observed many times as calibration sources by both XMM-Newton and Chandra.


\section{Observations and data analysis}

The dates, observing modes and exposure times of the observations are given in Table~\ref{modes}. The EPIC-pn data were processed using the XMM-Newton SAS (Science Analysis System) Version 5.4, using the \verb/epproc/ task, and spectra and lightcurves were extracted from a source region of 40'', using both single and double events, with background subtraction being performed using regions of the same size. Time intervals affected by proton flaring were filtered out of the event lists used for spectral extraction; the medium filter exposures of Markarian~501 were disregarded due to a high proton background throughout in both of them. Response matrices and ancillary response files were generated using the SAS tasks \verb/rmfgen/ and \verb/arfgen/ respectively. It was found, using \verb/epatplot/, that the minimum useful energy for spectral fitting is $\sim$0.4~keV. There are no pn data for PKS~0548-322.

The EPIC-MOS data were extracted with \verb/emproc/ under SAS V5.4. The imaging mode MOS data for all objects were found to be piled-up, and so were not used further in the analysis. Timing Uncompressed (TU) mode data were available for H1426+428 and Markarian~501; these did not suffer from pile-up, although the response matrices were found to have an incorrect energy scale in the region of the oxygen K-edge at $\sim$ 0.5~keV which biased the spectral fitting. Because of this, and the existence of better-quality pn data, the TU MOS data were not used in the final spectral fits. 

RGS1 and RGS2 data were extracted using \verb/rgsproc/ under SAS V5.4, which carries out background subtraction using regions spatially adjacent to the source. Time intervals affected by proton flaring were filtered out. For each source, RGS1 and RGS2 data (both first and second order spectra) were combined channel by channel to achieve the best possible signal to noise. The RGS spectra were binned into groups of three channels, approximately equivalent to the spectral resolution. Both the RGS and pn X-ray spectra were analysed using xspec, with the search for detailed spectral features being carried out in SPEX 2.00 (Kaastra et al. \cite{kaastra2002}). Since PKS~0548-322 has no usable EPIC data, the RGS was used to obtain an X-ray lightcurve with \verb/evselect/, performing background subtraction using the average background lightcurve obtained from regions above and below the source in the cross-dispersion direction.

   \begin{table*}
    
      \caption[]{Observation dates, X-ray instrument modes and exposure times for the BL Lac observations reported here.}
         \label{modes}  
   \centering
         \begin{tabular}{llllllr}
            \hline
            \noalign{\smallskip}
            Object & observation date & EPIC-MOS1$^{\mathrm{a}}$ & EPIC-MOS2$^{\mathrm{b}}$ & EPIC-pn$^{\mathrm{c}}$ & RGS1$^{\mathrm{d}}$ & RGS2$^{\mathrm{e}}$ \\
            \noalign{\smallskip}
            \hline
            \noalign{\smallskip}
            1H1219+301    & 2001-06-11 & 29304 (SW M) & 29304 (FF M) & 28523 (SW M) & 29854 & 29854  \\
            \noalign{\smallskip}
            \hline
            \noalign{\smallskip}
            H1426+428     & 2001-06-16 & 4044 (TU M)  & 4044 (TU M)  & 66239 (SW M) & 67442 & 67454    \\
                          &            & 60407 (LW M) & 60357 (FF M) &              &       &          \\
                          &            &            &            &            &       &        \\
                          &            &            &            &            &       &       \\
                          &            &            &            &            &       &        \\
            \noalign{\smallskip}
            \hline
            \noalign{\smallskip}
            Markarian~501 & 2002-07-12 & 6802 (TU T)  & 8699 (FF M)  & 6910 (SW T)  & 10232 & 10227  \\
                          &            & 1012 (TU M)  &              & 1720 (SW M)  &       &        \\
                          & 2002-07-14 & 7177 (TU M)  & 11877 (FF M) & 6910 (SW T)  & 12654 & 12647  \\
                          &            & 3857 (TU T)  &              & 4201 (SW M)  &       &        \\
            \noalign{\smallskip}
            \hline
            \noalign{\smallskip}
            PKS~0548-322   & 2001-10-03                & 47303 (LW M) & 47277 (FF M) & 0 (SW M) & 48152 & 48161   \\
            \noalign{\smallskip}
            \hline
         \end{tabular}
\begin{list}{}{}
\item[$^{\mathrm{a}}$] Exposure time in seconds for MOS1. The operating mode and filter are given in brackets (SW = Small Window, TU = Timing Uncompressed, LW = Large Window; M = Medium filter, T = Thick filter).
\item[$^{\mathrm{b}}$] Exposure time in seconds for MOS2. The operating mode and filter are given in brackets (FF = Full Frame, TU = Timing Uncompressed, LW = Large Window; M = Medium filter).
\item[$^{\mathrm{c}}$] Exposure time in seconds for pn. The operating mode and filter are given in brackets (SW = Small Window; M = Medium filter, T = Thick filter).
\item[$^{\mathrm{d}}$] Exposure time in seconds for RGS1; all exposures were in Spectroscopy mode.
\item[$^{\mathrm{e}}$] Exposure time in seconds for RGS2; all exposures were in Spectroscopy mode.
\end{list}
  \end{table*}


\section{Lightcurves}

\begin{figure*}
   \centering
   \includegraphics[width=16cm]{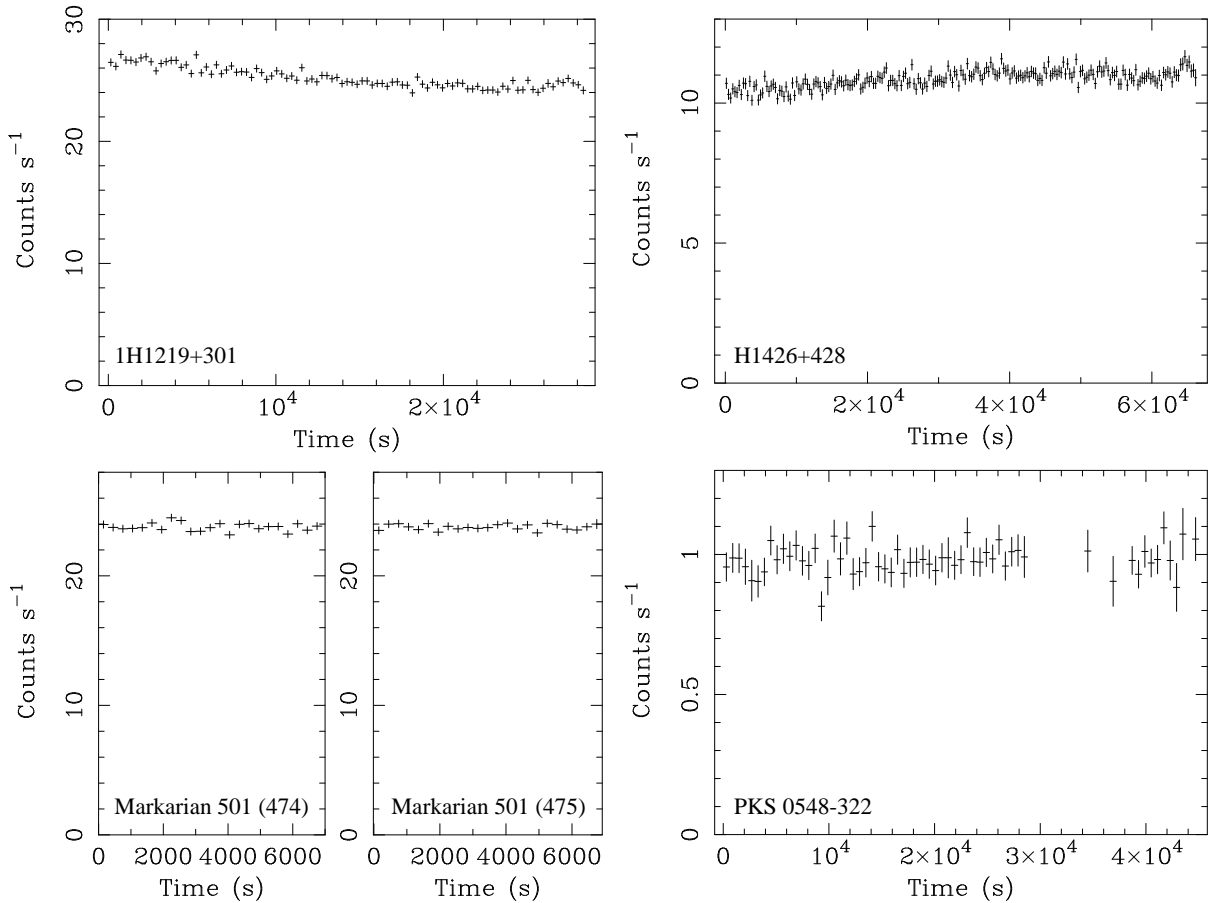}
      \caption{The 0.4$-$10~keV pn lightcurves of 1H1219+301, H1426+428 and Markarian~501 (from orbits 474 and 475), all in 300~s time bins, and the RGS lightcurve of PKS~0548-322 (0.35$-$2.5~keV; 600~s time bins).}
         \label{lc}
   \end{figure*}

The 0.4$-$10~keV pn lightcurves of 1H1219+301, H1426+428 and Markarian~501 are constant to within $\sim$ 5\%, and the RGS lightcurve of PKS~0548-322 is on average constant to within $\sim$ 10\% (the signal-to-noise is poorer for the RGS lightcurve); these lightcurves are plotted in Fig~\ref{lc}. Our spectral fitting is thus unlikely to be much affected by any spectral variability in the sources.

\section{EPIC-pn spectra}

\begin{figure*}
   \centering
   \includegraphics[width=16cm]{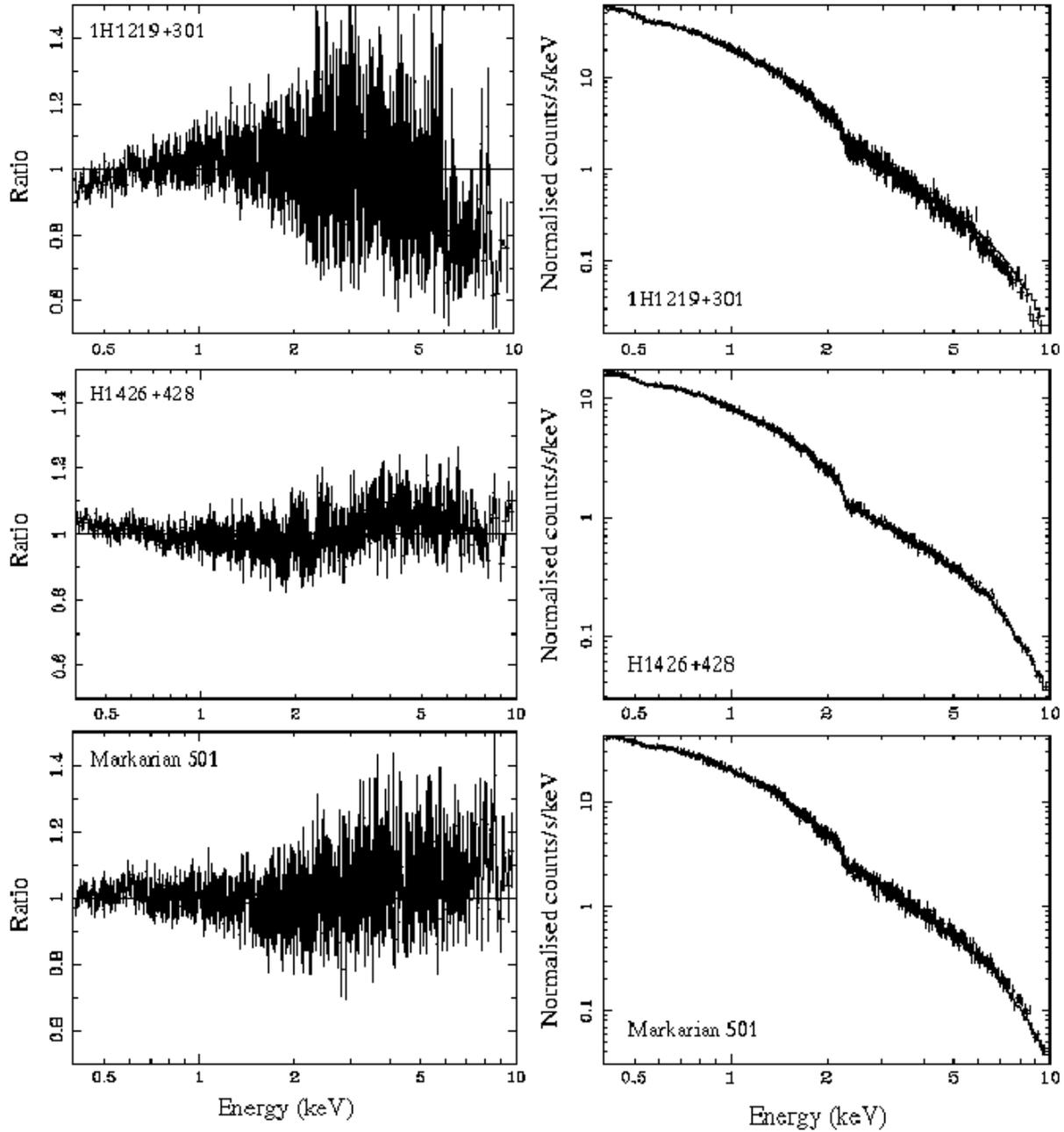}
      \caption{Left panels: ratios of the pn spectra of 1H1219+301, H1426+428 and Markarian~501 to fitted power-laws with fixed Galactic neutral absorption (plotted in the observed frame). Right panels: the pn spectra of 1H1219+301, H1426+428 and Markarian~501 with superimposed best-fit broken power-law models (with fixed Galactic neutral absorption; plotted in the observed frame). In all plots, the spectra are grouped to a minimum of 40, 350 and 100 counts per bin respectively.}
         \label{pn_spec}
   \end{figure*}

The 0.4$-$10 keV pn spectra were fitted with power-law models with fixed neutral absorption due to our Galaxy. The ratios of the data to these models are shown in Fig.~\ref{pn_spec}. Some spectral curvature is present in each case, but there is no evidence for deep warm absorption - or excess neutral absorption - in any of the sources; any fit residuals in the soft band are all within the 10\% calibration uncertainties which exist below 2~keV. Motivated by the observed spectral curvature, we fitted broken power-law models (again with fixed Galactic neutral absorption) to the spectra. The parameters of both broken power-law and simple power-law fits are listed in Table~\ref{specfits}, and the broken power-law fits are shown superimposed on the data in Fig.~\ref{pn_spec}.

   \begin{table*}
    
       \caption[]{Fit parameters for powerlaw and neutral absorption, and broken power-law with neutral absorption, fits to the EPIC-pn (0.4$-$10~keV) and RGS (0.35$-$2.5~keV) spectra of the BL Lacs objects observed. The 2$-$10~keV flux is given for the preferred model for each object.}
         \label{specfits}
   \centering
         \begin{tabular}{p{0.75in}p{0.6in}p{0.3in}p{0.7in}p{0.6in}p{0.7in}p{0.7in}p{0.6in}p{0.6in}}
            \hline
            \noalign{\smallskip}
            Object & Instrument & N$_{\rm H}$$^{\mathrm{a}}$ & $\Gamma$$_{\rm 1}$$^{\mathrm{b}}$ & E$_{\rm{B}}$$^{\mathrm{c}}$ & $\Gamma$$_{\rm 2}$$^{\mathrm{d}}$ & N$^{\mathrm{e}}$ & $\chi$$^{\rm 2}_{\rm red}$$^{\mathrm{f}}$ & F$_{\rm 2-10}$$^{\mathrm{g}}$ \\
            \noalign{\smallskip}
            \hline
            \noalign{\smallskip}
1H1219+301    & pn  & 1.78$^{\mathrm{h}}$ & 2.485 $\pm$ 0.006 &                 &                   & 2.23 $\pm$ 0.08   & 1.38 (714)  &                 \\ 
              & pn  & 1.78$^{\mathrm{h}}$ & 2.40 $\pm$ 0.02   & 1.4 $\pm$ 0.2   & 2.61 $\pm$ 0.02   & 2.29 $\pm$ 0.02   & 0.99 (712)  & 2.64 $\pm$ 0.02 \\
              & RGS & 1.78$^{\mathrm{h}}$ & 2.09 $\pm$ 0.02   &                 &                   & 2.00 $\pm$ 0.02   & 1.08 (1038) &                 \\
              & RGS & 1.78$^{\mathrm{h}}$ & 1.76 $\pm$ 0.06   & 0.75 $\pm$ 0.05 & 2.44 $\pm$ 0.07   & 2.5  $\pm$ 0.1    & 0.89 (1036) &                 \\
            \hline
H1426+428     & pn  & 1.36$^{\mathrm{i}}$ & 1.870 $\pm$ 0.003 &                 &                   & 0.899 $\pm$ 0.002 & 1.61 (607)  &                 \\
              & pn  & 1.36$^{\mathrm{i}}$ & 1.922 $\pm$ 0.009 & 1.5 $\pm$ 0.2   & 1.816 $\pm$ 0.009 & 0.891 $\pm$ 0.003 & 1.24 (605)  & 2.90 $\pm$ 0.01 \\
              & RGS & 1.36$^{\mathrm{i}}$ & 1.68  $\pm$ 0.02  &                 &                   & 0.778 $\pm$ 0.006 & 1.19 (996)  &                 \\
              & RGS & 1.36$^{\mathrm{i}}$ & 1.47  $\pm$ 0.07  & 0.79 $\pm$ 0.08 & 1.88 $\pm$ 0.05   & 0.89  $\pm$ 0.05  & 1.05 (994)  &                 \\
            \hline
Markarian~501 & pn  & 1.73$^{\mathrm{h}}$ & 2.308 $\pm$ 0.005 &                 &                   & 2.639 $\pm$ 0.008 & 1.17 (653)  &                 \\
              & pn  & 1.73$^{\mathrm{h}}$ & 2.34  $\pm$ 0.01  & 1.9 $\pm$ 0.4   & 2.23 $\pm$ 0.03   & 2.62  $\pm$ 0.01  & 1.03 (651)  & 4.48 $\pm$ 0.02 \\
              & RGS & 1.73$^{\mathrm{h}}$ & 2.00  $\pm$ 0.02  &                 &                   & 2.27  $\pm$ 0.02  & 1.01 (1038) &                 \\
              & RGS & 1.73$^{\mathrm{h}}$ & 1.70  $\pm$ 0.06  & 0.71 $\pm$ 0.04 & 2.26 $\pm$ 0.05   & 2.8   $\pm$ 0.1   & 0.84 (1036) &                 \\
            \hline
PKS~0548-322  & RGS & 2.51$^{\mathrm{i}}$ & 1.74  $\pm$ 0.02  &                 &                   & 0.942 $\pm$ 0.008 & 1.28 (1038) &                 \\
              & RGS & 2.51$^{\mathrm{i}}$ & 1.38  $\pm$ 0.05  & 0.74 $\pm$ 0.02 & 2.01 $\pm$ 0.04   & 1.19  $\pm$ 0.04  & 1.03 (1036) & 2.47 $\pm$ 0.08 \\
            \noalign{\smallskip}
            \hline
         \end{tabular}
\begin{list}{}{}
\item[$^{\mathrm{a}}$] neutral absorbing column in 10$^{\rm 20}$ cm$^{\rm -2}$
\item[$^{\mathrm{b}}$] slope of soft power-law
\item[$^{\mathrm{c}}$] break energy in keV
\item[$^{\mathrm{d}}$] slope of hard power-law
\item[$^{\mathrm{e}}$] power-law normalisation in 10$^{\rm -2}$ photons keV$^{\rm -1}$ cm$^{\rm -2}$ s$^{\rm -1}$ at 1 keV
\item[$^{\mathrm{f}}$] reduced $\chi$$^{\rm 2}$ of the total fit; the degrees of freedom are given in brackets
\item[$^{\mathrm{g}}$] 2$-$10~keV flux of preferred model in 10$^{\rm -11}$ erg cm$^{\rm -2}$ s$^{\rm -1}$
\item[$^{\mathrm{h}}$] fixed at Galactic Hydrogen column (Elvis et al. \cite{elvis})
\item[$^{\mathrm{i}}$] fixed at Galactic Hydrogen column (Costamante et al. \cite{costamante})
\end{list}
  \end{table*}

\section{RGS spectra}

\begin{figure*}
   \centering
   \includegraphics[width=18cm]{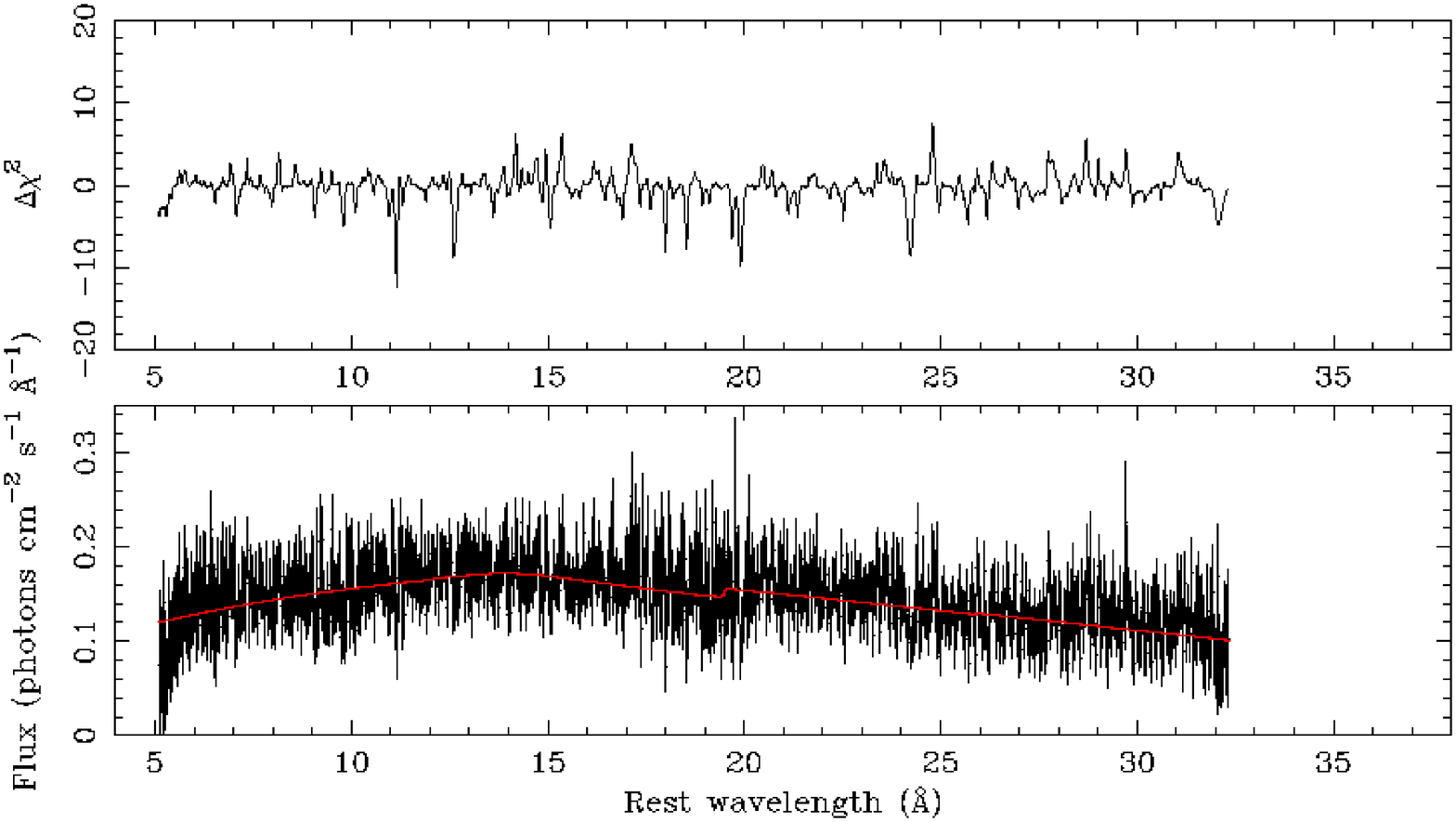}
      \caption{The statistical significance of narrow absorption and emission features in the RGS spectra of 1H1219+301 (top), and the RGS spectrum itself with best-fit broken power-law model (red) superimposed (bottom), plotted in the source rest frame. 
              }
         \label{1h_rgs}
   \end{figure*}

\begin{figure*}
   \centering
   \includegraphics[width=18cm]{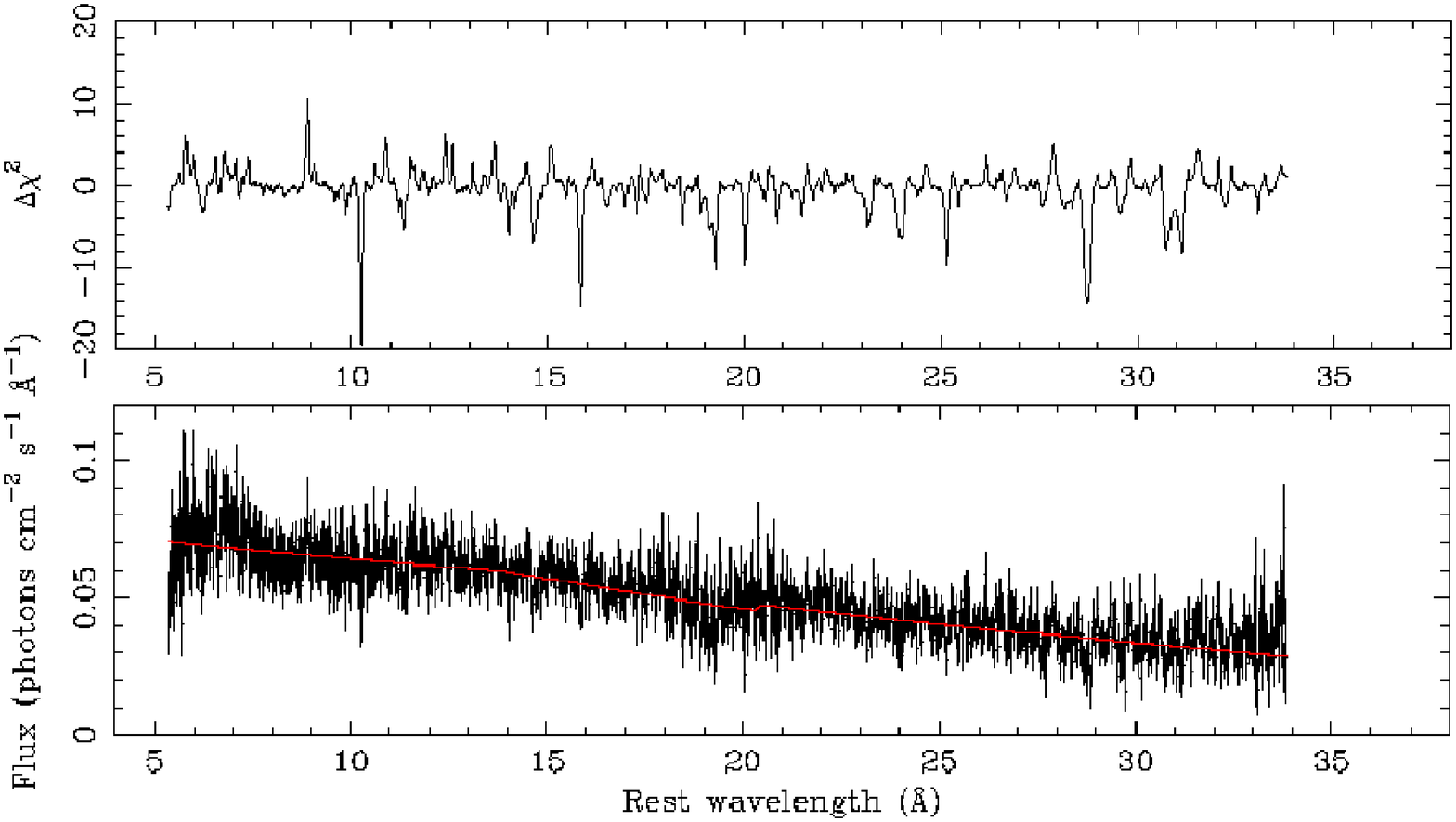}
      \caption{The statistical significance of narrow absorption and emission features in the RGS spectra of H1426+428 (top), and the RGS spectrum itself with best-fit broken power-law model (red) superimposed (bottom), plotted in the source rest frame. 
              }
         \label{h1_rgs}
   \end{figure*}

\begin{figure*}
   \centering
   \includegraphics[width=18cm]{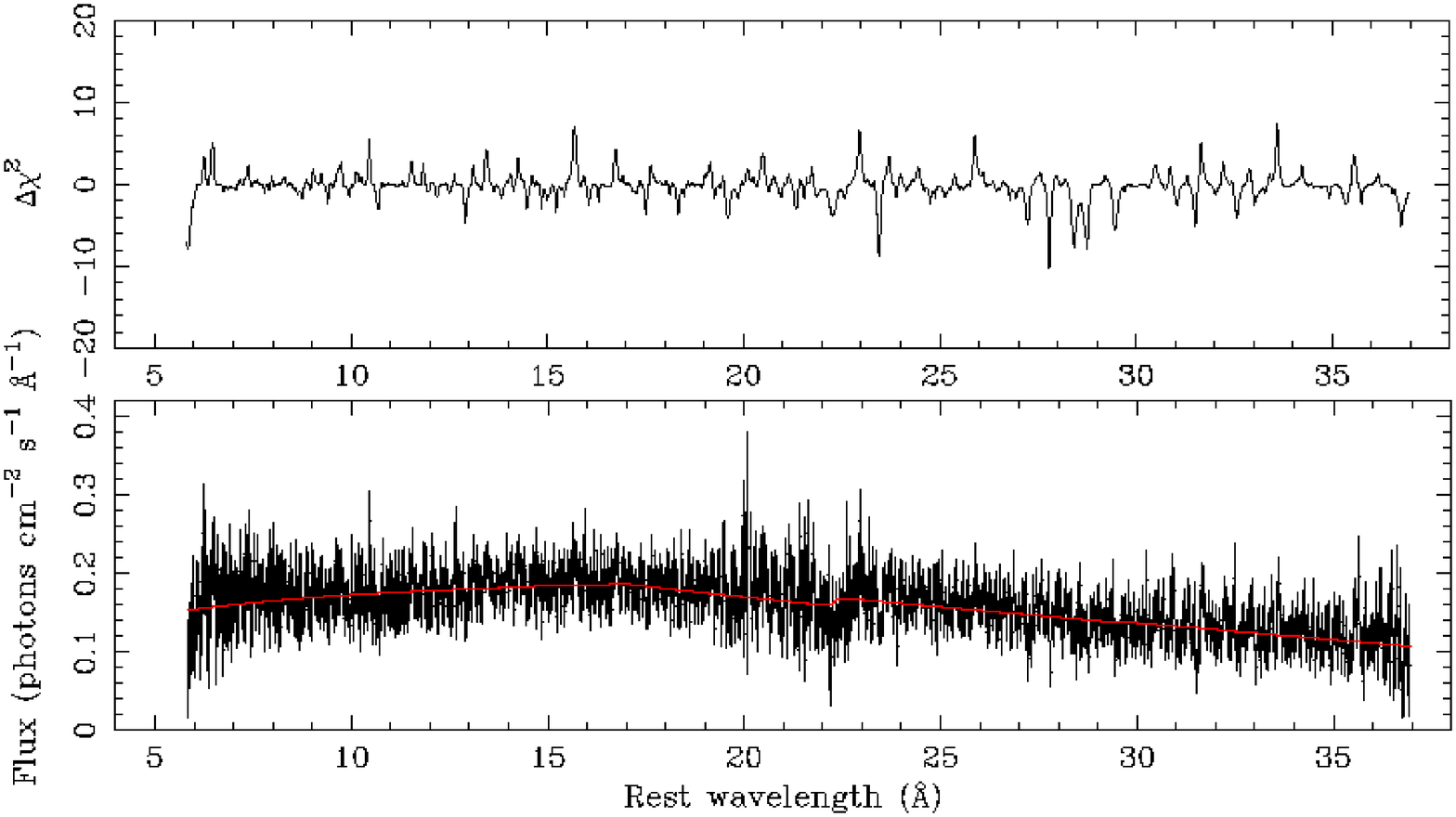}
      \caption{The statistical significance of narrow absorption and emission features in the RGS spectra of Markarian~501 (top), and the RGS spectrum itself with best-fit broken power-law model (red) superimposed (bottom), plotted in the source rest frame. 
              }
         \label{mkn501_rgs}
   \end{figure*}

\begin{figure*}
   \centering
   \includegraphics[width=18cm]{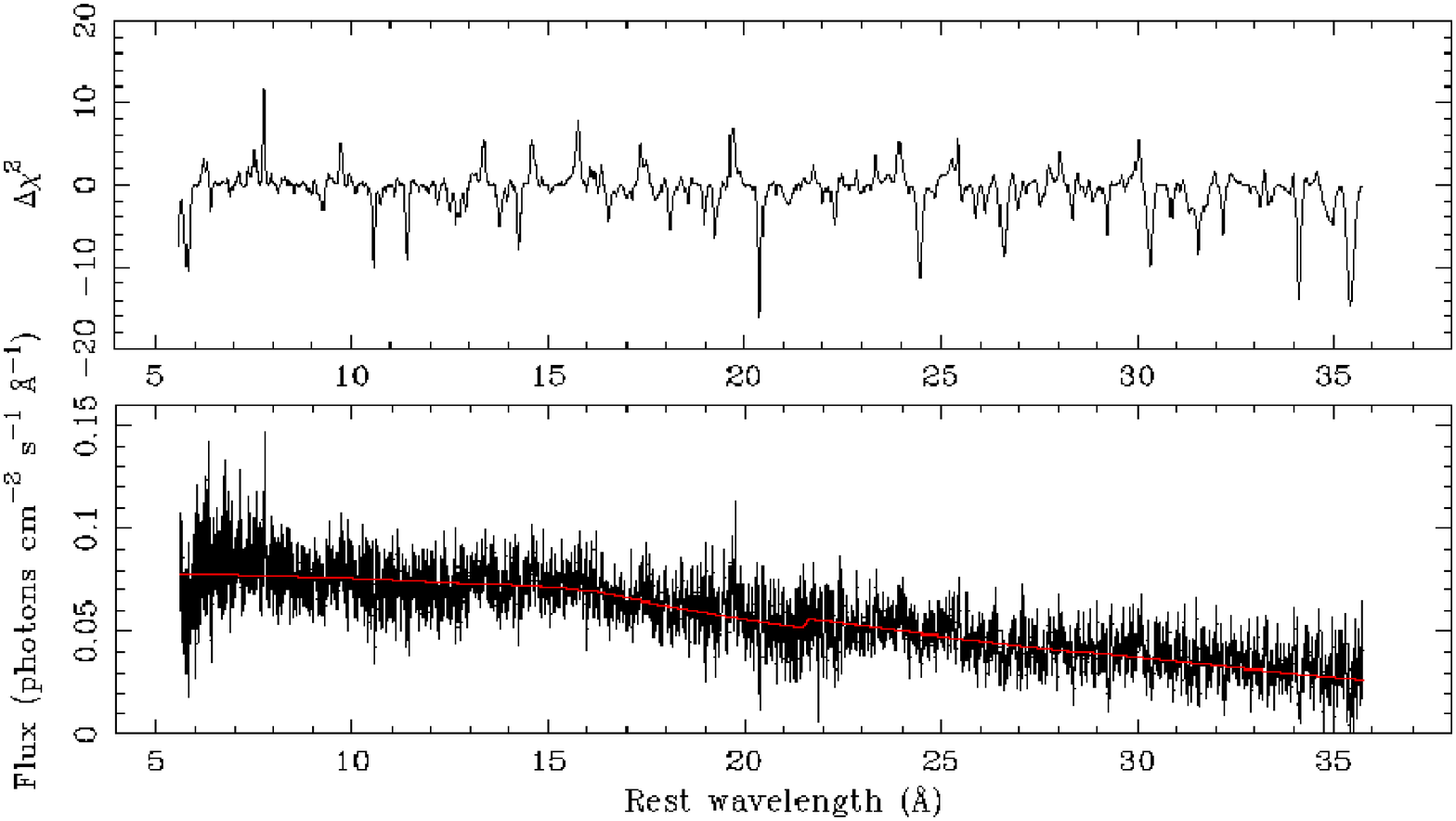}
      \caption{The statistical significance of narrow absorption and emission features in the RGS spectra of PKS~0548-322 (top), and the RGS spectrum itself with best-fit broken power-law model (red) superimposed (bottom), plotted in the source rest frame. 
              }
         \label{pks0548_rgs}
   \end{figure*}

None of the RGS spectra show evidence of broad absorption features. The parameters of power-law and broken power-law continuum fits, with neutral absorption due to the ISM of our Galaxy, are given in Table~\ref{specfits}. The statistical significance of apparent narrow emission and absorption features in the spectra was assessed by fitting a gaussian against a local continuum (as done for NGC~7469 in Blustin et al. \cite{blustin}). The results of this, alongside the respective RGS spectra, are plotted in Figs.~\ref{1h_rgs}--\ref{pks0548_rgs}. If the features were purely due to statistical noise, then in an RGS spectrum of about 1000 data points, one would expect to see about three features significant at 3$\sigma$. Any 4$\sigma$ features would be likely to be real.

Figs.~\ref{1h_rgs} and \ref{mkn501_rgs} show that the apparent features in 1H1219+301 and Markarian~501 are consistent with statistical noise. H1426+428 (Fig.~\ref{h1_rgs}) and PKS~0548-322 (Fig.~\ref{pks0548_rgs}) each have one narrow absorption feature significant at 4$\sigma$ or greater. The H1426+428 feature has an equivalent width of 40 $\pm$ 10~m$\rm \AA$ at 11.55~$\rm \AA$ (observed frame), which is 10.23~$\rm \AA$ in the source rest frame, and could be identified as \ion{Ne}{x} Ly$\beta$. The corresponding Ly$\alpha$ is not present though, making the identification unlikely. The observed line could conceivably be \ion{Ne}{x} Ly$\alpha$ blueshifted by $\sim$ 47000 km s$^{\rm -1}$, if it was associated with the BL Lac jet, but no further lines are found which could give support to this pattern. The 4$\sigma$ absorption line in PKS~0548-322 appears at 21.82~$\rm \AA$ in the observed frame (20.41~$\rm \AA$ in the rest frame), and has an equivalent width of 70 $\pm$ 20~m$\rm \AA$ at 21.82~$\rm \AA$. It has no obvious interpretation in the rest frame - the nearest likely transition is \ion{O}{vii} He$\alpha$, at a blueshift of $\sim$ 17000 km s$^{\rm -1}$. Again, there are no further features to reinforce this interpretation.

H1426+428 and PKS~0548-322, respectively, have five and ten features significant at three sigma. In each case, only one of the features is in emission. This is not what is expected from a random distribution, and suggests that the features may be real. They are not due to bad channels in the spectrum, and the vast majority do not seem to be related to known systematic errors (e.g. anomalies in the response matrix at CCD gaps). We have, however, been unable to find any consistent or plausible identification or interpretation for any of the features.



\section{Discussion and conclusions}

Our high resolution and CCD resolution spectra showed no sign of deep broad ionised absorption features, as had been claimed to be present in spectra from previous missions. Two RGS spectra, those of PKS~0548-322 and H1426+428, have weak indications of narrow line absorption, although we are not able to find convincing identifications for any of these apparent features, and so cannot confirm that they are real. 

\begin{figure*}
   \centering
   \includegraphics[width=17cm]{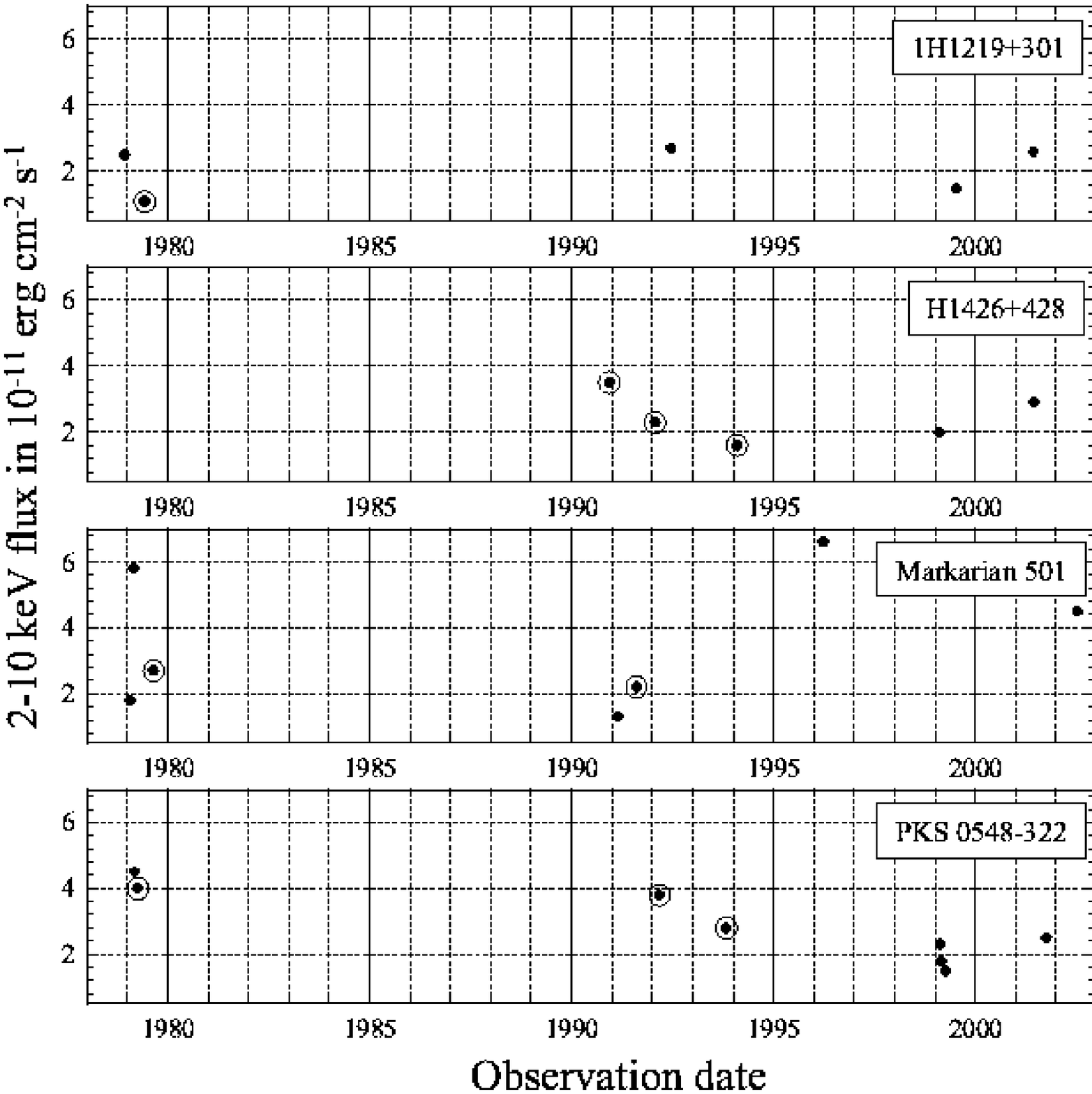}
      \caption{2$-$10~keV fluxes of 1H1219+301, H1426+428, Markarian~501 and PKS~0548-322 observed with Einstein, ROSAT, BBXRT, ASCA, BeppoSAX and XMM-Newton. Observations where intrinsic absorption was reported are circled.
              }
         \label{bllac_features_history}
   \end{figure*}

So where have the deep intrinsic absorption features gone? Fig.~\ref{bllac_features_history} contains plots of the observed 2$-$10~keV X-ray flux versus time, over the past 25 years, of the four sources discussed in this paper. Those observations where features have been reported are marked. Only published observations by missions that had sufficient soft X-ray effective area and spectral resolution to detect the features are included (Einstein, ROSAT, BBXRT, ASCA, BeppoSAX and XMM-Newton; there have been no Chandra publications on these objects). It has been claimed (e.g. Madejski et al. \cite{madejski1991}) that intrinsic absorption could be detected in EXOSAT spectra, by means of a soft X-ray spectral flattening, even though this mission had no energy resolution in the band where the features occur; we find no reports in the literature that EXOSAT has ever actually detected such absorption for any of our four objects, so EXOSAT observations are not included. ROSAT's energy resolution in the relevant energy band is extremely low, but since ROSAT spectra of H1426+428 and Markarian~501 have shown possible evidence of ionised absorption (Sambruna et al. \cite{sambruna1997_h1}, Sambruna et al. \cite{sambruna1998_pks0548}, Fink et al. \cite{fink}), we include flux measurements from this mission. The fluxes, when not quoted directly in the 2$-$10~keV band in the original papers, were calculated from the original fit parameters using either PIMMS or XSPEC, and the derived values are given in Table~\ref{bllac_fluxes} alongside the observation dates and references. The 2$-$10~keV band was chosen as it is the one quoted most frequently in the literature (thus minimising the chance of introducing errors in the extrapolation to a new band), and also the range least affected by the possible confusion between power-law slope and excess neutral absorption in the soft band.

   \begin{table*}
    
       \caption[]{The history of the 2$-$10~keV fluxes of 1H1219+301, H1426+428, Markarian~501 and PKS~0548-322 as observed by past X-ray missions.}
         \label{bllac_fluxes}
   \centering
         \begin{tabular}{p{1in}p{1in}p{0.8in}p{0.7in}p{0.6in}p{1.7in}}
            \hline
            \noalign{\smallskip}
Object & Mission & Date & F$_{\rm (2-10)}$$^{\mathrm{a}}$ & Intrinsic absorption & Reference for flux value\\
            \noalign{\smallskip}
            \hline
            \noalign{\smallskip}
1H1219+301 & XMM-Newton & 11/06/01 & 2.6 & no & Table~\ref{specfits} \\
           & BeppoSAX   & 13/07/99 & 1.5 & no & Costamante et al. \cite{costamante} \\
           & ROSAT      & 18/06/92 & 2.7 & no & Lamer et al. \cite{lamer} \\
           & Einstein   & 05/06/79 & 1.1 & yes$^{\mathrm{b}}$ & Urry et al. \cite{urry} \\
           & Einstein   & 07/12/78 & 2.5 & no & Urry et al. \cite{urry} \\
            \hline								
H1426+428  & XMM-Newton & 16/06/01 & 2.9 & no & Table~\ref{specfits} \\
           & BeppoSAX   & 08/02/99 & 2.0 & no & Costamante et al. \cite{costamante} \\
           & ASCA       & 06/02/94 & 1.6 & yes & Sambruna et al. \cite{sambruna1997_h1} \\
           & ROSAT      & 24/01/92 & 2.3 & yes & Sambruna et al. \cite{sambruna1997_h1} \\
           & BBXRT      & 09/12/90 & 3.5 & yes & Sambruna et al. \cite{sambruna1997_h1} \\
            \hline								
Markarian~501 & XMM    & 13/07/02 & 4.5 & no & Table~\ref{specfits} \\
           & ASCA      & 26/03/96 & 6.6 & no & Kubo et al. \cite{kubo} \\
           & ROSAT     & 13/08/91 & 2.2 & yes & Fink et al. \cite{fink} \\
           & ROSAT     & 25/02/91 & 1.3 & no & Lamer et al. \cite{lamer} \\
           & Einstein  & 22/08/79 & 2.7 & yes & Ciliegi et al. \cite{ciliegi} \\
           & Einstein  & 01/03/79 & 5.8 & no & Ciliegi et al. \cite{ciliegi} \\
           & Einstein  & 25/01/79 & 1.8 & no & Ciliegi et al. \cite{ciliegi} \\
            \hline								
PKS~0548-322  & XMM      & 03/10/01 & 2.5 & no & Table~\ref{specfits} \\
             & BeppoSAX  & 07/04/99 & 1.5 & no & Costamante et al. \cite{costamante} \\
             & BeppoSAX  & 26/02/99 & 1.8 & no & Costamante et al. \cite{costamante} \\
             & BeppoSAX  & 20/02/99 & 2.3 & no & Costamante et al. \cite{costamante} \\
             & ASCA      & 30/10/93 & 2.8 & yes & Sambruna \& Mushotzky \cite{sambruna1998_pks0548} \\
             & ROSAT     & 06/03/92 & 3.8 & yes$^{\mathrm{c}}$ & Lamer et al. \cite{lamer} \\
             & Einstein  & 06/04/79 & 4.0 & yes & Sambruna \& Mushotzky \cite{sambruna1998_pks0548} \\
             & Einstein  & 10/03/79 & 4.5 & no & Ciliegi et al. \cite{ciliegi} \\
            \noalign{\smallskip}
            \hline
         \end{tabular}
\begin{list}{}{}
\item[$^{\mathrm{a}}$] 2$-$10~keV flux in 10$^{\rm -11}$ erg cm$^{\rm -2}$ s$^{\rm -1}$
\item[$^{\mathrm{b}}$] Madejski et al. \cite{madejski1991}
\item[$^{\mathrm{c}}$] Sambruna \& Mushotzky \cite{sambruna1998_pks0548}
\end{list}
  \end{table*}

The data in Fig.~\ref{bllac_features_history} suggest that the features - if they exist - are only visible some of the time. However, absorption could have been present but not reported (or not detectable) during some of these observations, especially the earlier ones. Because the XMM-Newton spectra are of much higher quality than any of the previous data, and no broad features are observed in any of the XMM-Newton spectra, we can state categorically that if the features observed with past missions were real, they must be a transient phenomenon.

In Fig.~\ref{bllac_features_history} no clear patterns emerge in terms of feature visibility at different flux levels. In H1426+428, a feature of some sort was observed for at least three years, but is no longer seen eight years after it was first detected. In Markarian~501, the absorption appears in a space of a few months and has disappeared by the time of the next observation four years hence. We can use the detection rate of the broad absorption features in previous observations to determine the probability of observing no features with XMM-Newton. The fraction of previous observations which did not yield a detection of a broad absorption feature is 75\%, 25\%, 66\% and 57\% for 1H1219+301, H1426+428, Markarian~501 and PKS~0548-322 respectively. With these detection rates, the probability of observing no broad absorption features in any of these objects with XMM-Newton is only 7\%.

If we take these results at face value, then we can rule out the existence of transient deep, broad intrinsic absorption features in these sources at 93\% confidence. If a further comparable set of observations fail to detect any features, then their existence will be ruled out at 99.5\% confidence. Time - and future observations with the high quality instruments on board XMM-Newton and Chandra - will lay the question of broad absorption features in BL Lacs to rest.

Low spectral resolution, low signal-to-noise and calibration uncertainties must have contributed to previous conclusions about these features. The major source of the apparent detection of intrinsic soft X-ray absorption features was, though, probably the use of a single power-law to fit the X-ray spectral continuum of BL Lac objects. These continua were fitted at higher energies and extrapolated over the soft band, and the resulting deficit of photons in the soft band was interpreted as evidence for absorption. As our observations show, there is significant spectral curvature in our four sources. This is also seen in other recent observations; for example, Fossati et al. (\cite{fossati}) have shown that the X-ray continuum of Markarian~421 is best described by a gradually steepening power-law with the photon index increasing towards higher energies. Such spectral curvature, when fitted with a simple power-law, would give rise to soft X-ray residuals which could plausibly be interpreted as absorption.

So, BL Lac objects do not seem likely to yield much knowledge of the immediate environment of their central engines through X-ray absorption features. A more fruitful search for such evidence will probably involve other classes of radio galaxy in which we are not looking directly down the axis of the jet; Broad Line Radio Galaxies with Seyfert-like nuclei are an ideal candidate for such studies.


\begin{acknowledgements}
      This work is based on observations obtained with XMM-Newton, an ESA science mission with instruments and contributions directly funded by ESA Member States and the USA (NASA). We acknowledge the support of PPARC. We wish to thank the referee, Grzegorz Madejski, for helpful comments on the contribution of continuum fitting to the detection of absorption features in data from previous missions. This research has made use of the NASA/IPAC Extragalactic Database (NED) which is operated by the Jet Propulsion Laboratory, California Institute of Technology, under contract with the National Aeronautics and Space Administration.
\end{acknowledgements}

\end{document}